\documentclass[twocolumn, tighten]{aastex631}

\shorttitle{GS-EXAM-I: GJ~1151 }

\begin{document}

\title{uGMRT Survey of EXoplanets Around M-dwarfs (GS-EXAM): Radio observations of GJ~1151}

\author[0000-0002-0554-1151]{Mayank Narang}
\affiliation{Academia Sinica Institute of Astronomy \& Astrophysics, \\ 11F of Astro-Math Bldg., No.1, Sec. 4, Roosevelt Rd., Taipei 10617, Taiwan}
\affiliation{Department of Astronomy \& Astrophysics, Tata Institute of Fundamental Research,\\
Homi Bhabha Road, Colaba,
Mumbai, 400005, India}

\author[0000-0002-3530-304X]{Manoj Puravankara}
\affiliation{Department of Astronomy \& Astrophysics, Tata Institute of Fundamental Research,\\
Homi Bhabha Road, Colaba,
Mumbai, 400005, India}

\author[0000-0002-0872-181X]{H. K. Vedantham}
\affiliation{ASTRON, Netherlands Institute for Radio Astronomy, Oude Hoogeveensedijk 4, Dwingeloo, 7991 PD, The
Netherlands}

\author{C. H. Ishwara Chandra}
\affiliation{National Centre for Radio Astrophysics, TIFR,\\ Post Bag 3, Ganeshkhind, Pune 411007, India}

\author[0000-0001-8845-184X]{Ayanabha De}
\affiliation{Department of Astronomy \& Astrophysics, Tata Institute of Fundamental Research,\\
Homi Bhabha Road, Colaba,
Mumbai, 400005, India}

\author[0000-0002-9497-8856]{Himanshu Tyagi}
\affiliation{Department of Astronomy \& Astrophysics, Tata Institute of Fundamental Research,\\
Homi Bhabha Road, Colaba,
Mumbai, 400005, India}

\author[0000-0001-8075-3819]{Bihan Banerjee}
\affiliation{Department of Astronomy \& Astrophysics, Tata Institute of Fundamental Research,\\
Homi Bhabha Road, Colaba,
Mumbai, 400005, India}

\author[0000-0002-4638-1035]{Prasanta K. Nayak}
\affiliation{Instituto de Astrofísica, Pontificia Universidad Católica de Chile, Av. Vicuña MacKenna 4860, 7820436, Santiago, Chile}

\author[0000-0002-9967-0391]{Arun Surya}
\affiliation{Indian Institute of Astrophysics, 2nd block Koramangala, Bangalore, 560034, India}

\author[0000-0002-2585-0111]{B. Shridharan}
\affiliation{Department of Astronomy \& Astrophysics, Tata Institute of Fundamental Research,\\
Homi Bhabha Road, Colaba,
Mumbai, 400005, India}

\author{Vinod C. Pathak}
\affiliation{Department of Astronomy \& Astrophysics, Tata Institute of Fundamental Research,\\
Homi Bhabha Road, Colaba,
Mumbai, 400005, India}

\author[0009-0007-2723-0315]{Mihir Tripathi}
\affiliation{Academia Sinica Institute of Astronomy \& Astrophysics, \\ 11F of Astro-Math Bldg., No.1, Sec. 4, Roosevelt Rd., Taipei 10617, Taiwan}

\begin{abstract}

{Coherent radio emission with properties similar to planetary auroral signals has been reported from GJ 1151, a quiescent, slow-rotating mid-M star, by the LOFAR Two-metre (120-170~MHz) Sky Survey (LoTSS). The observed {LOFAR} emission is fairly bright at 0.89 mJy with 64\% circular polarization, and the emission characteristics are consistent with the interaction between an Earth-sized planet with an orbital period of 1-5 days and the magnetic field of the host star. However, no short-period planet has been detected around GJ 1151. To confirm the reported radio emission caused by the putative planet around GJ 1151 and to investigate the nature of this emission, we carried out uGMRT observations of GJ~1151  at 150, 218, and 400 MHz over 33 hours across ten epochs.  No emission was detected at any frequency. {While at 150~MHz and 218 MHz, non-detection could be due to the low sensitivity of our observations, at 400 MHz, the rms sensitivities achieved were sufficient to detect the emission observed with LOFAR at $\sim$~20$\sigma$ level}. Our findings suggest that the radio emission is highly time-variable, likely influenced by the star-planet system's phase and the host star's magnetic field. Additional observations below 170~MHz, at more frequent epochs (as the periodicity of the emission is unknown), especially during periods of high stellar magnetic field strength, are needed to confirm the emission.}

\end{abstract}

\section{Introduction} \label{Intro}
Since the initial detection of radio emission originating from Jupiter by \cite{burke55}, astronomers have pondered the possibility of utilizing radio observations of stars as a possible method for detecting exoplanets \citep{Fennelly74, Yantis77, Winglee86b}. Consequently, numerous studies have been conducted searching for radio emissions from exoplanets due to star-planet interactions (SPI){ \citep[e.g.,][]{Bastian00, Etangs09, Lazio10, Etangs11, Hallinan13, Etangs13, Bastian18, Lynch18, Narang21b, Narang20, Turner20, prez20, 2022MNRAS.515.2015N,2023NatAs...7..569P, 2023A&A...670A.124C, 2024MNRAS.528.2136S,2024MNRAS.tmp..579N,2024arXiv240316392T}, }yet no conclusive detection has been reported so far.

Recently \cite{Vedantham20} used data from the LOFAR Two-metre Sky Survey (LoTSS) to detect radio emission originating from an M dwarf star, GJ 1151. The observations were conducted between 120 - 170 MHz over an 8-hour period. The time-frequency averaged Stokes I and V flux densities from the star are 0.89~mJy and 0.57~mJy, respectively. The radio emission is highly polarized with a circularly polarized fraction of 64 $\pm$ 6\%. Additionally, the radio emission is flat and does not vary with frequency between 120 - 170 MHz. Similar plateaued emission is observed from planets in our solar system \cite{1992AdSpR..12h..99Z}. 

The GJ~1151 system was observed three more times by LoTSS (for 8 hours each epoch), but no emission was detected from the position of the star for the other three epochs. The time variability of the emission and high circular polarization are inconsistent with extragalactic sources.  Based on these facts, and in combination with the positional coincidence with GJ~1151, \cite{Vedantham20} concluded that the emission they detected with LOFAR is from the star and is not some background source. 

GJ~1151 is categorized as an M4.5 dwarf, located 8 pc away from Earth \citep{Gaia21}. The star has  high proper motion, with pmRA = 1.545\arcsec~yr$^{-1}$ and pmDe = -0.962\arcsec~yr$^{-1}$. Considering this high proper motion, GJ~1151 would have traversed around 12\arcsec.3 between the initial LOFAR detection (J2014.454) and our observations conducted in March 2021. This notable positional difference can help distinguish any background radio source as a potential emitter. 

Follow-up studies to characterize GJ~1151 indicate very little chromospheric activity, suggesting a relatively quiescent star. The radio emission detected by LOFAR does not conform to the Gudel Benz relation \citep{2021ApJ...919L..10P,CC}: the X-ray emission observed from GJ~1151 is roughly three orders of magnitude lower than predicted based on the Gudel Benz relation \citep{1993ApJ...405L..63G}. Given the star's quiescent nature and the highly polarized, time-variable emission, a plausible hypothesis for the observed radio emission is the interaction between the star and an undetected planet.  \cite{Vedantham20} suggested that the radio emission may represent a scaled-up version of the emissions observed in the Jupiter-Io coupling phenomenon \citep{Zarka07,saur13, Turnpenney}, where the host star's magnetic field drives the emission, rather than the planet itself.

The observed LOFAR emission from GJ 1151 closely resembles the theoretical predictions from the interaction between an M-type host star and an Earth-sized planet orbiting within a period between 1-5 days \citep{Vedantham20}. However, subsequent radial velocity (RV) follow-up studies conducted by \cite{Pope20}, \cite{Per21}, and \cite{2023A&A...671A..50B} \citep[also see][]{Suv21}, were unable to confirm the presence of such a planet. Nonetheless, these studies placed strict upper limits on the planet's mass, restricting it to $\leq 1.2 M_\oplus$, consistent with the emission models employed in \cite{Vedantham20}. Similarly transit observations of the system by \cite{2021ApJ...919L..10P} also failed to detect any signatures of a planet in close orbit.  

{\cite{2023A&A...671A..50B} reported the detection of a long-period planet with a 390-day orbit and a mass  M$_P > 10.6$ M$_\oplus$. The expected flux density due to SPI with this long-period planet is estimated to be far below 0.1 mJy \citep{2023A&A...671A..50B}, which is too faint to detect. The same study also presented tentative evidence for another sub-stellar companion or long-period magnetic variability originating from the star. However, these observations were inconclusive in ruling out the possibility of a low-mass, short-period planet that could be the source of the observed radio emission.
It remains possible that the radio emission is caused by the interaction between one of these long-period planets and an exomoon, or some other source of local plasma, similar to the interaction between Jupiter and Io (see also Section 4). }

In a recent study, \citet{CC} detected coherent radio emission from 19 M dwarfs utilizing LoTSS data. Their results imply that coherent radio emission is prevalent among main sequence M dwarfs.  It is plausible that a portion of the observed radio emission may arise from plasma emission. However, the emission originating from quiescent, slow-rotating stars cannot be solely explained by plasma emission, making them potential systems to explore for signatures of star-planet interaction (SPI).

Furthermore, radio emissions have also been detected from M dwarf stars known to host exoplanets, such as Proxima Centauri \citep{prez20} and YZ Ceti \citep{2023arXiv230500809T}. These detections exhibit a high degree of circular polarization and demonstrate modulations corresponding to the orbital phase of the planet. However, neither these modulations nor the detection of GJ1151 constitute conclusive proof for SPI, as there remains the possibility that electrons accelerated in stellar flares could have generated the observed ECMI emission \citep{2024NatAs...8...50Y}. Follow-up radio observations aimed at building sufficient statistical evidence to demonstrate modulation at the synodic period of the planet are necessary.

We observed GJ~1151 using the upgraded Giant Metrewave Radio Telescope (uGMRT) as part of the \textit{uGMRT Survey of EXoplanets Around M-dwarfs (GS-EXAM) program}. The GS-EXAM program aims at observing all planet-hosting M dwarfs and field brown dwarfs within 20 pc. This is one of the largest uGMRT programs that focuses on observing radio emission from stellar and sub-stellar object and has been awarded more than 300 hrs of uGMRT time.

The previous study by \cite{Vedantham20} using LOFAR had only detected the emission in one 8-hour epoch out of the four pointings, indicating temporal variability. Moreover, the emission detected by LOFAR fell within the frequency range of 120-170 MHz. Hence, to explore and quantify the emission's variability and investigate its nature at higher frequencies ($>$170 MHz), we observed the GJ~1151 system across different frequency bands: 150 MHz, 218 MHz (using band 2 of uGMRT, encompassing the range of 120-250 MHz), and 400 MHz (in band 3, spanning 250-500 MHz). Section 2 provides details of the observations and data reduction. The results of our observations and discussion are presented in Sections 3 and 4, respectively, while the summary is provided in Section 5.

\begin{table}[]
\centering
\caption{Summary and log of observation and the rms sensitivity reached during our observation run.}
\label{Table1}
\footnotesize
\begin{tabular}{cccc}
\\\hline
Date & Start time& Obs time& rms \\  
 &  (UTC) &  (hr) & (mJy/beam) \\  \hline

\multicolumn{4}{c}{150 MHz} \\ \hline
2021 March 18 & 1330 & 4 & 2.45 \\
2021 March 19 & 1330 & 4 & 2.51 \\
2021 March 21 & 1330 & 4 & 6.32 \\
2021 March 22 & 1330 & 4 & 4.79 \\
2021 March 23 & 1330 & 4 & 2.25 \\ \hline
\multicolumn{4}{c}{218 MHz} \\ \hline
2021 March 18 & 1330 & 4 & 0.39 \\
2021 March 19 & 1330 & 4 & 0.38 \\
2021 March 21 & 1330 & 4 & 0.82 \\
2021 March 22 & 1330 & 4 & 0.67 \\
2021 March 23 & 1330 & 4 & 0.43 \\ \hline
\multicolumn{4}{c}{400 MHz} \\ \hline
2021 March 18 & 1830 & 3 & 0.04 \\
2021 March 19 & 1830 & 3 & 0.04 \\
2021 March 21 & 1830 & 3 & 0.05 \\
2021 March 22 & 2030 & 2 & 0.05 \\
2021 March 23 & 1830 & 2 & 0.05\\
\hline
\end{tabular}%
\label{T1}
\end{table}

\section{Observation and data reduction}
The GJ~1151 system was observed for 33 hrs with uGMRT (proposal ID 39\_008, PI Mayank Narang). We observed the system at two uGMRT bands, band 2 (120-250 MHz) and band 3 (250-500 MHz). Since the presence of the planet is not confirmed, and its properties are unknown, we can not observe the system at specific phases similar to  \cite{prez20,2023arXiv230500809T}. However, based on the modeling of the radio emission from GJ~1151, it has been suggested that the planet's orbital period could fall within the range of one to five days \citep{2020NatAs.tmp...34V}. In light of the radio emission being variable,  we decided to conduct observations of the GJ~1151 system for nearly five consecutive days. This strategy aims to cover the broadest possible range of the orbital phases of the putative planet. To ensure near-simultaneous detection in both band 2 and band 3, the observations in the two bands were conducted sequentially, one after the other. Great care was taken to synchronize the start and end times of each observation run across all five days, thereby maximizing coverage across the orbital phase spectrum for any potential planet. 

In band 2 (120-250 MHz), observations of GJ~1151 were conducted from March 18, 2021, to March 23, 2021, spanning UTC 13:30 to 17:30 each day (refer to Table \ref{T1} for the log of observations). The primary flux calibrators, 3C147 and 3C286, were observed at the start (3C147) and end (3C286) of the observations, both observed for 5 minutes each. For phase calibration, 1021+219 was used, operating in a loop with the science target GJ~1151, with a 30-minute integration on GJ~1151 and a 5-minute integration on 1021+219.

The observations at 400 MHz (band 3; 250-500 MHz) were also conducted from March 18, 2021, to March 23, 2021. However, due to the operational constraints of the telescope, the observations did not start immediately after the band 2 observations. Nevertheless, they were conducted within 1-3 hours after the termination of the band 2 observations. Additionally, three out of the five observations lasted for 3 hours each, while the remaining two observations spanned 2 hours each. For calibration, 3C286 served as both the primary flux calibrator and the phase calibrator. 3C286 was observed in a loop with GJ~1151, with 5 minutes on 3C286 and 30 minutes on GJ~1151.  {Since only Band~4 (550-900 MHz) of uGMRT has been tested and calibrated for polarization observation \citep{2020arXiv200408542D,2023arXiv230504420K}, we have not carried out the polarisation analysis for our target.  }

The band 3 (250-500 MHz) data was reduced using the standard uGMRT data reduction pipeline, called \textit{CASA} Pipeline-cum-Toolkit for Upgraded Giant Metrewave Radio Telescope data REduction \citep[CAPTURE;][]{CAPTURE}. We also applied a primary beam correction to the image to account for the drop in the flux away from the phase center.

The CAPTURE pipeline, however, is not suited for reducing band 2 uGMRT data. To reduce the band 2 data, we used Source Peeling and Atmospheric Modeling (SPAM) pipeline \citep{Intema09b, Intema14b, Intema14}. SPAM is a Python-based pipeline that was developed to reduce low-frequency radio interferometric observations.  The SPAM pipeline however does not support processing large fractional bandwidths ($\delta f/f>~0.2$). Therefore natively, the SPAM is not capable of reducing the wideband data from uGMRT. However, based on the nature of band-2 (120-250 MHz) with a break in the band in the middle, we have decided to divide the band~2 observations into two frequency ranges with a bandwidth of about $\sim$ 30 MHz around regions of relatively low RFI \citep[also see][]{2023MNRAS.522.1662N}.  These smaller chunks (sub-bands)  can be processed independently. We selected channel numbers from 600-1100 (500 channels) corresponding to a band center of $\sim$ 218 MHz and channel numbers from 1350-1750 (400 channels) corresponding to a band center of $\sim$ 150 MHz. These channels were relatively free of RFI and were used to produce the final images.

\begin{figure*}
\centering
\includegraphics[width=0.5\linewidth]{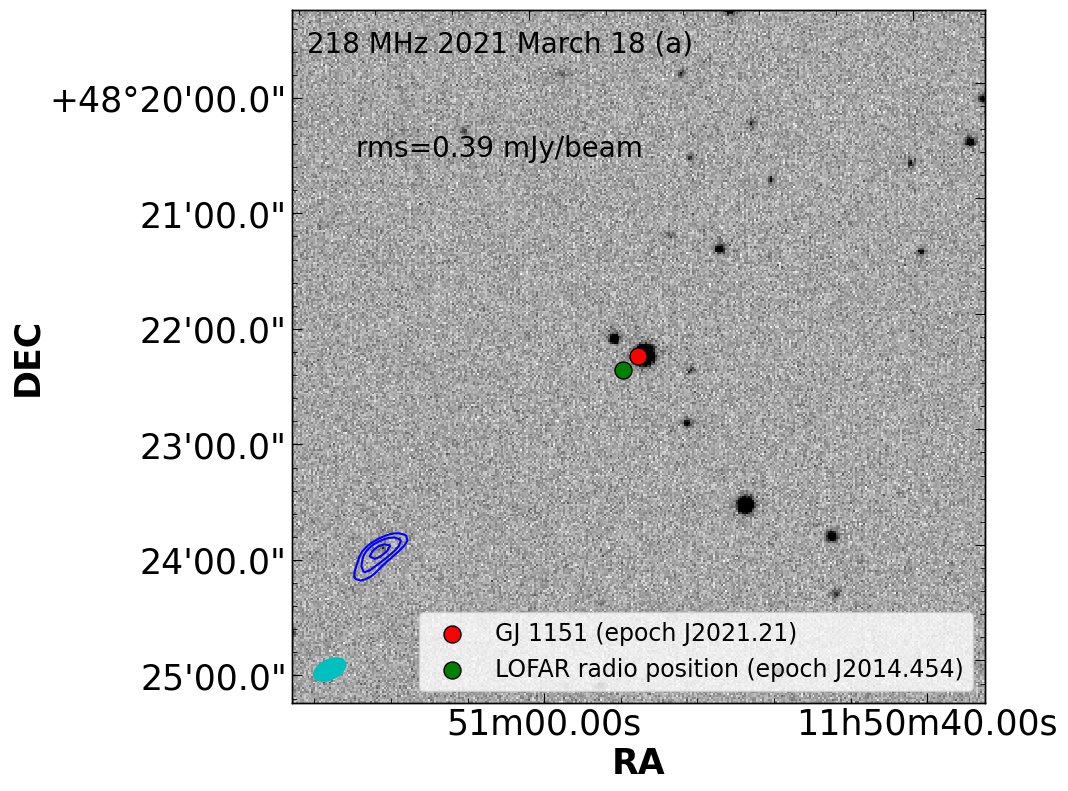}\includegraphics[width=0.5\linewidth]{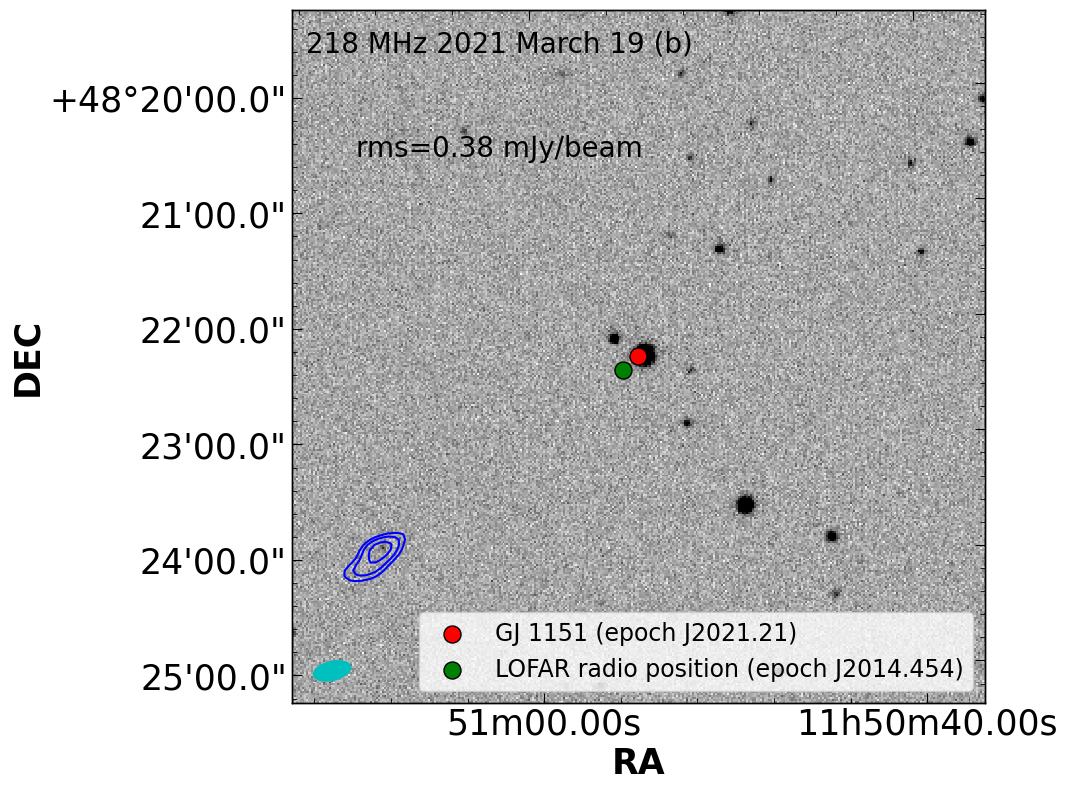}

\includegraphics[width=0.5\linewidth]{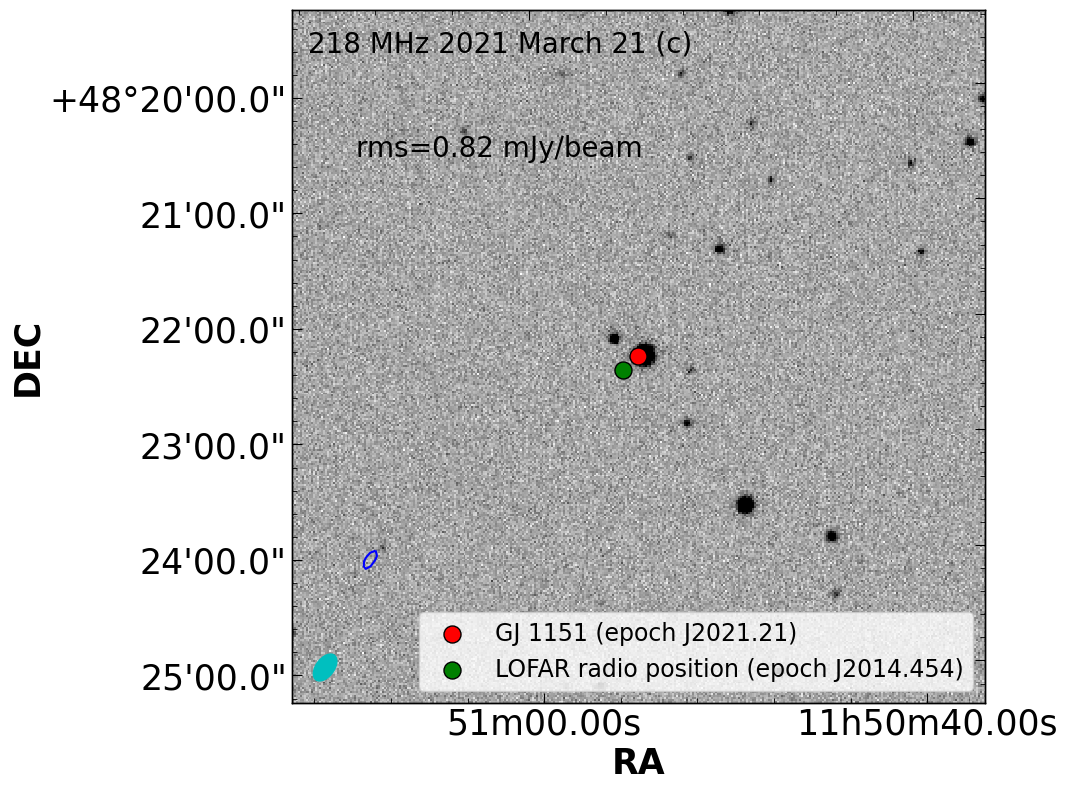}\includegraphics[width=0.5\linewidth]{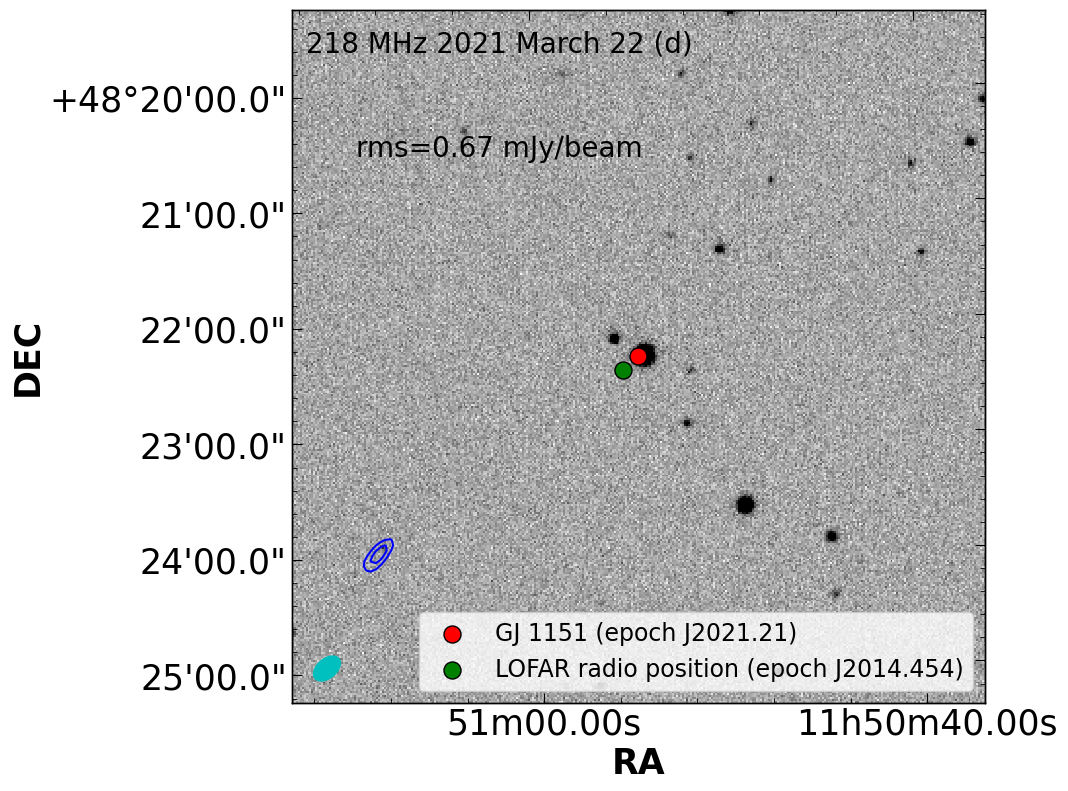}

\includegraphics[width=0.5\linewidth]{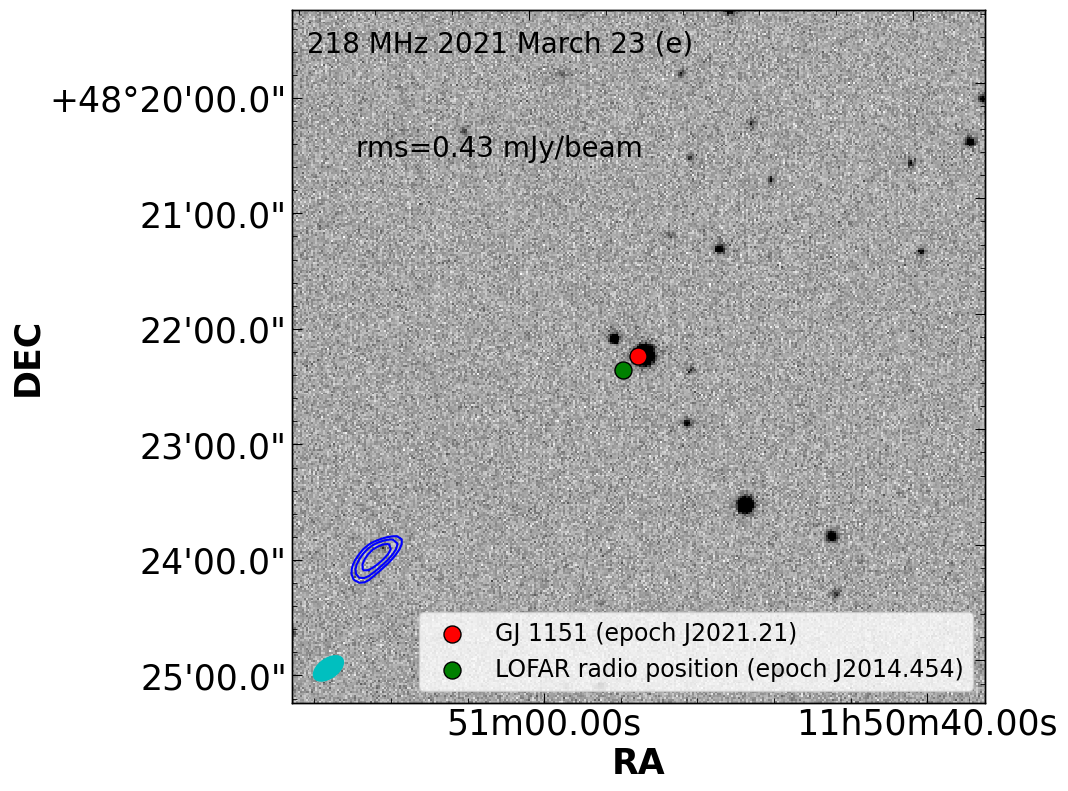}
\caption{The uGMRT image (blue contours) of the GJ~1151 field at 218~MHz overlaid on the ZTF zg band image (epoch J2021.17). The red circle marks the position of GJ~1151 at the time of observations (epoch J2021.21), and the green circle marks the position of the radio source detected with LOFAR {(epoch J2014.454)}. The rms achieved for each of the observations are mentioned in the top left corner. The contours plotted are   5, 7, and 10~$\times\;\sigma$. The beam is shown as a cyan ellipse at the bottom left corner. }
\label{fig4b}
\end{figure*}

\begin{figure*}
\centering
\includegraphics[width=0.5\linewidth]{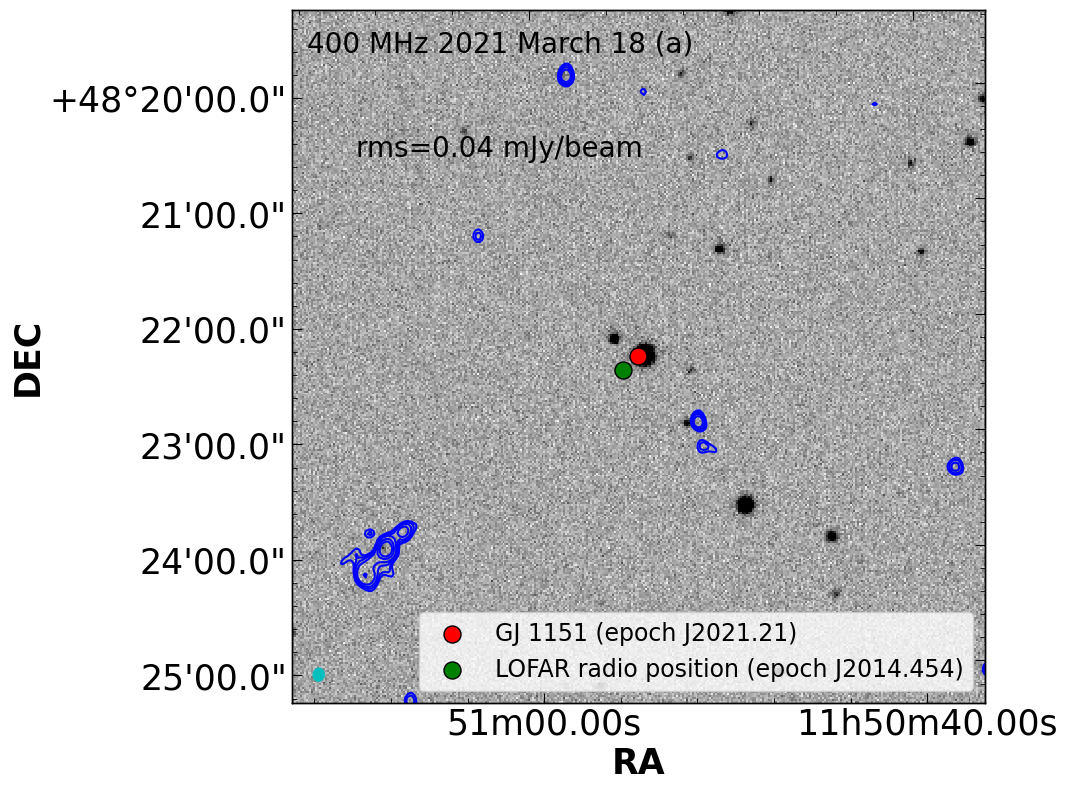}\includegraphics[width=0.5\linewidth]{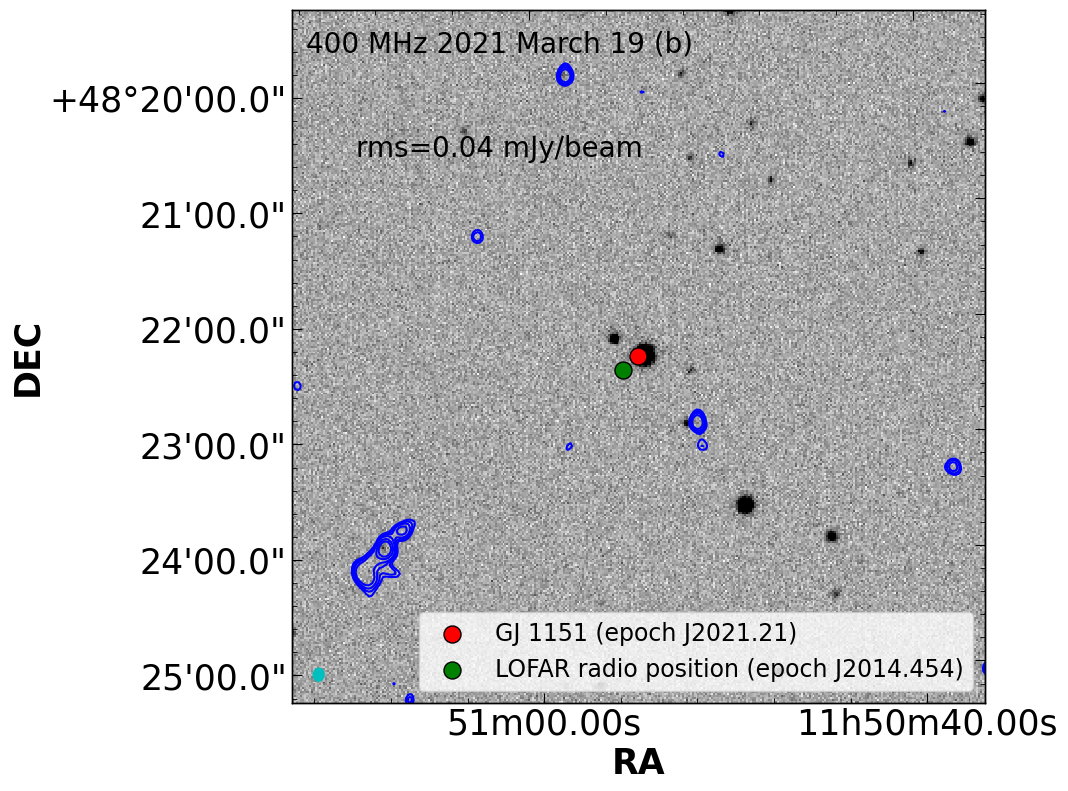}

\includegraphics[width=0.5\linewidth]{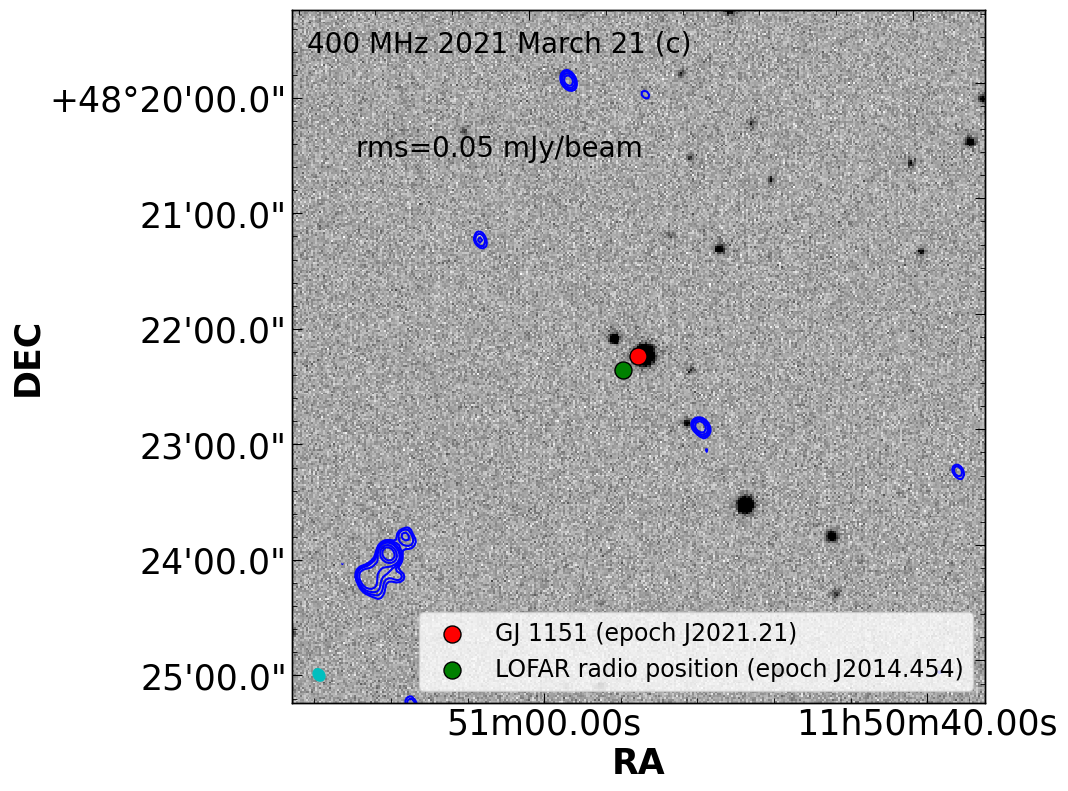}\includegraphics[width=0.5\linewidth]{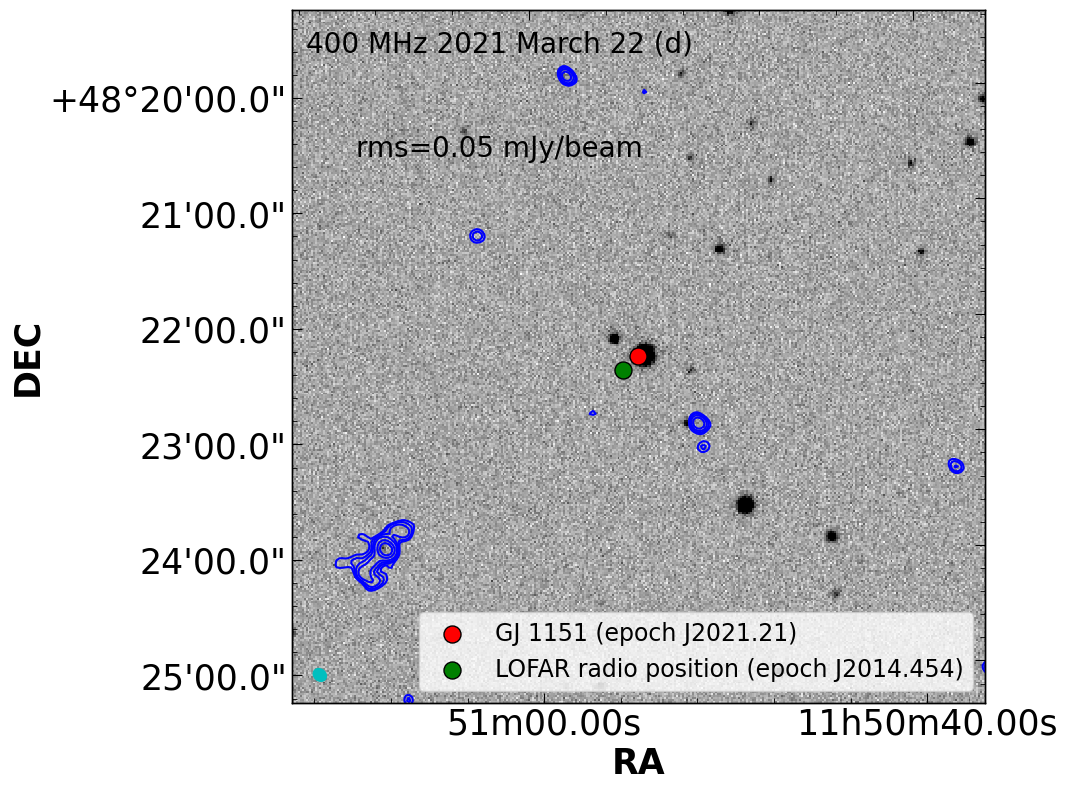}

\includegraphics[width=0.5\linewidth]{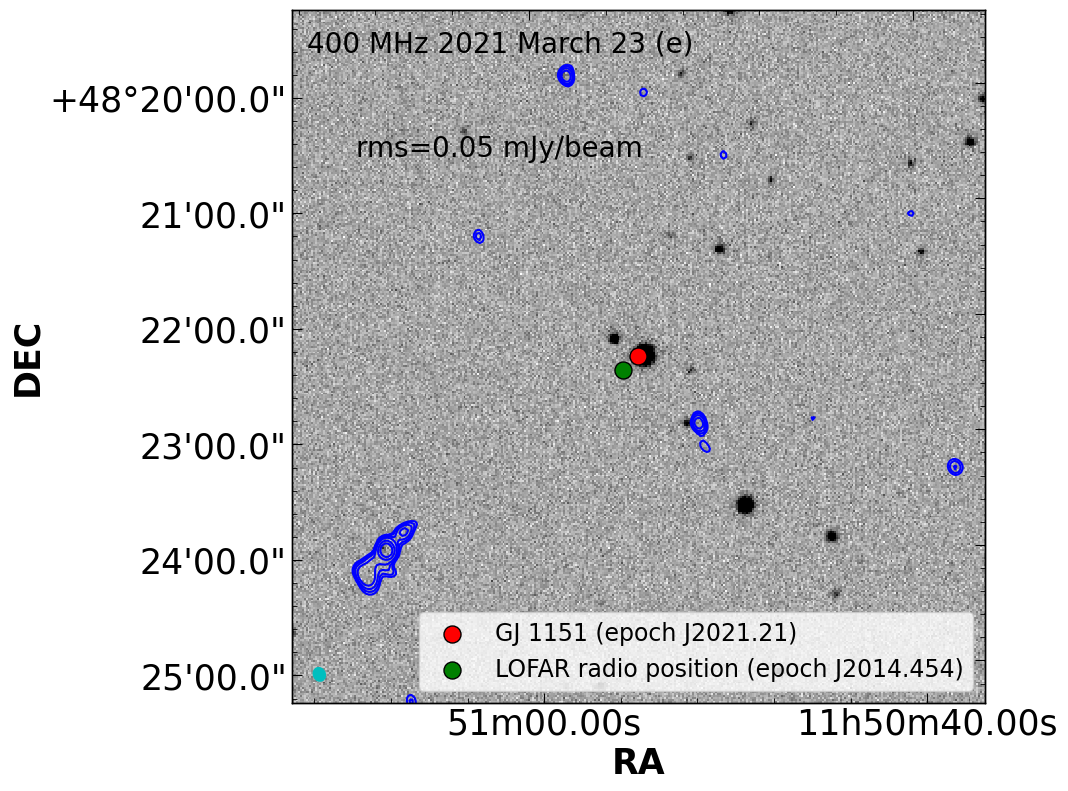}
\caption{The uGMRT image (blue contours) of the GJ~1151 field at  400~MHz overlaid on the ZTF zg band image (epoch J2021.17). The red circle marks the position of GJ~1151 at the time of observations (epoch J2021.21), and the green circle marks the position of the radio source detected with LOFAR {(epoch J2014.454)}. The rms achieved for each of the observations are mentioned in the top left corner. The contours plotted are  5, 7, 10, 15, 30, and 50~$\times\;\sigma$. The beam is shown as a cyan ellipse at the bottom left corner. }
\label{fig4c}
\end{figure*}

\section{Result}

The GJ~1151 system was observed with uGMRT for five almost consecutive days in band 2  (120-250 MHz) and band 3 (250-500 MHz), totalling 33 hrs. In Figure \ref{fig4b}, the band 2 218 MHz images obtained on all 5 days. We have overlaid the uGMRT image (blue contours) on top of the ZTF zg band image. The ZTF image was taken on 2021 March 02. This was the closest epoch optical image to our radio observations. {No radio source was detected within $\sim$ 6\arcmin of GJ 1151 at 150 MHz, hence we do not show any images at that frequency.}

In Table \ref{T1}, we have listed the rms achieved for each of the nights. In band 2, at  150 MHz, we achieve an rms sensitivity between 2.25-6.32 mJy/beam. This is well above the detected flux of 0.89 mJy with LOFAR \citep{Vedantham20}. No radio emission was detected from the source either at 150 MHz or 218 MHz. The rms sensitivity at 218 MHz (band 2) is between 0.38-0.67 mJy/beam. 
If the emission at $\sim$218 MHz from the source were similar to the emission seen between 120-170 MHz, the rms sensitivities achieved in two out of the five runs were low enough for at least a 2 $\sigma$ detection.

The images of the GJ~1151 field at 400 MHz (band~3) are shown in Figure \ref{fig4c}.  At 400 MHz, the rms sensitivity we achieved is 0.04-0.05 mJy/beam (see Table \ref{T1}). This is about 18-22 times smaller than the flux detected by LOFAR. Despite the deep observations, no radio emission was detected from the source on any of the nights at 400 MHz.

\section{Discussion}



The radio emission detected from Proxima Centauri by \cite{prez20} and YZ Ceti by \cite{2023arXiv230500809T} also seem to exhibit modulation corresponding to the orbital period of their respective planets. Since the period for the putative planet around GJ~1151 is not known, it is possible that all ten of the pointings were carried out at a phase where no emission is expected from the system. Several factors such as the inclination of the planet's orbit around the star, the exact orbital period, and the emission cone opening angle, are unknown. This makes knowing the exact configuration of the star-planet system when the emission was detected by \cite{Vedantham20} almost impossible. 

Furthermore, the Zeeman-Doppler imaging (ZDI) maps of GJ 1151 \citep{2024MNRAS.527.4330L} reveal that the stellar magnetic obliquity shows large-scale variations. The results from \cite{2024MNRAS.527.4330L}, indicate that during our observations GJ 1151 was in a magnetically quiescent state. They determined the average unsigned dipole magnetic field for the host star GJ 1151 to range between 26 and 62 Gauss. Consequently, the emission peaks around $\sim$ 64-176 MHz. As a result, the flux at higher frequencies diminishes significantly, potentially rendering our observations at 400 MHz too high to detect the emission from the system. It is possible that because of these reasons we could have missed the detection during our observation run.

\cite{2023A&A...671A..50B} found that the star shows episodes of higher magnetic activity, which induces variability in RVs, activity indices, and photometry.  One possibility is that the emission is not due to SPI but produced entirely by the star. However, as argued in \cite{Vedantham20}, the large duration ($>$ 8 hrs), high degree of polarization ($> 64 \%$ circular polarization), and broad-band emission are similar to planetary emission from the solar system, and not stellar.

The expected radio emission due to SPI with GJ~1151 c (with a mass of $>$ 10.6 M$_{\oplus}$ and period of $\sim$390 days) is $<<$ 0.1 mJy (at $\sim$140 MHz) \citep{2023A&A...671A..50B}. Therefore we can rule out the emission from GJ~1151 c.  Another plausible explanation is that the emission originates from the planet GJ~1151 c itself. The emission frequency of $\sim$ 140 MHz suggests that the magnetic field strength of the planet must be equal to or greater than 50 G (with the cyclotron frequency $\nu_c$ = 2.8 $\times$ B[G], where B[G] represents the magnetic field strength in Gauss). \cite{2023AJ....165....1N} have shown that for a magnetic field strength of $\sim$ 50~G, the planet mass has to be $\sim$~1~$M_J$, much larger than the estimated mass of  GJ~1151 c. 

{\cite{2023A&A...671A..50B} also suggested the presence of a long-period substellar companion to the star. Given that radio emissions from brown dwarfs are commonly detected \citep[e.g.,][]{2006ApJ...648..629B, 2008ApJ...676.1307B, 2008ApJ...684..644H, 2013A&A...549A.131A, 2016ApJ...830...85R}, it is possible that this substellar companion could be the source of the observed emission. To investigate the existence of such a companion around GJ 1151, a JWST Cycle 3 program  (PID 5497; \citealt{2024jwst.prop.5497S}) has been approved. This program will use JWST/NIRCam coronagraphy with the F200W and F444W filters, which are sensitive enough to detect companions with masses as low as 3 M$_J$ at distances of 3 to 25 au.}

\section{Summary}

The system GJ~1151 was observed for 33 hrs over ten epochs with uGMRT (at 150 MHz, 218 MHz, and 400 MHz) for five almost consecutive days. However, no radio emission was detected towards the source. The rms sensitivity we achieved at 150 MHz is much larger than the detected flux with LOFAR from the system; hence no detection was possible at 150 MHz.  If the flux level at $\sim$ 200 MHz is similar to that at 150 MHz, then we should have detected the emission for at least two out of the five pointings in band 2 at 218 MHz. At 400 MHz we achieved an rms between 0.04-0.05 mJy. These are about 18-22 times lower than the observed flux from the system. However, given that the unsigned dipole magnetic field for the star is between 26 - 62 G, the peak emission frequency is between $\sim$ 64-176 MHz. Consequently, the flux at higher frequencies is much smaller than what was observed with LOFAR. Since the magnetic field of the star is time variable and the exact phase of the SPI is unknown,{ additional observations below 170~MHz, at more frequent epochs (as the periodicity of the emission is unknown), especially during periods of high stellar magnetic field strength, are needed  to better understand the nature of the emission, and place strong constraints on the properties of the putative plane}.

\section{Data availability}

{The data presented in this article are available on the GMRT archive. The GMRT data can be accessed from  https://naps.ncra.tifr.res.in/goa/,  with proposal id $39\_008$. In addition, the reduced fits files are available on request from the corresponding author.}

\section{Acknowledgement}
{This work is based on observations made with the Giant Metrewave Radio Telescope, operated by the National Centre for Radio Astrophysics of the Tata Institute of Fundamental Research, and is located at Khodad, Maharashtra, India. We thank the GMRT staff for their efficient support of these observations. In addition, we acknowledge the support of the Department of Atomic Energy, Government of India, under Project Identification No. RTI4002. This research has also used NASA's Astrophysics Data System Abstract Service and the SIMBAD database, operated at CDS, Strasbourg, France. }

\bibliographystyle{aasjournal}

\bibliography{ms}
\end{document}